\documentclass[twocolumn,preprintnumbers,amsmath,aps]{revtex4}
\usepackage{graphicx}
\usepackage{dcolumn}
\usepackage{bm}
\usepackage{color}
\begin{document}
\title{Characterization of the superconducting phase in tellurium hydride at high pressure} 
\author{Tomasz P. Zem{\l}a$^{1}$}
\author{Klaudia M. Szcz{\c e}{\' s}niak$^{2}$}
\author{Adam Z. Kaczmarek$^{3*}$}
\author{Svitlana V. Turchuk$^{1,4}$}
\affiliation{$^1$ Institute of Physics, Jan D{\l}ugosz University in Cz{\c{e}}stochowa, Ave. Armii Krajowej 13/15, 42-200 Cz{\c{e}}stochowa, Poland}
\affiliation{$^2$ Faculty of Chemistry, University of Warsaw, Pasteura 1 Str., 02-093 Warsaw, Poland}
\affiliation{$^3$ Institute of Physics, Cz{\c{e}}stochowa University of Technology, Ave. Armii Krajowej 19, 42-200 Cz{\c{e}}stochowa, Poland}
\affiliation{$^4$ Department of Physics, Lesya Ukrainka East European National University, Ave. Volya 13, 43000 Lutsk, Ukraine}
\email{akaczmarsky1410@gmail.com}

\begin{abstract}

At present, hydrogen-based compounds constitute one of the most promising classes of materials for applications as a phonon-mediated high-temperature superconductors. Herein, the behavior of the superconducting phase in tellurium hydride (HTe) at high pressure ($p=300$ GPa) is analyzed in details, by using the isotropic Migdal-Eliashberg equations. The chosen pressure conditions are considered here as a case study which corresponds to the highest critical temperature value ($T_{c}$) in the analyzed material, as determined within recent density functional theory simulations. It is found that the Migdal-Eliashberg formalism, which constitutes a strong-coupling generalization of the Bardeen-Cooper-Schrieffer (BCS) theory, predicts that the critical temperature value ($T_{c}=52.73$ K) is higher than previous estimates of the McMillan formula. Further investigations show that the characteristic dimensionless ratios for the the thermodynamic critical field, the specific heat for the superconducting state, and the superconducting band gap exceeds the limits of the BCS theory. In this context, also the effective electron mass is not equal to the bare electron mass as provided by the BCS theory. On the basis of these findings it is predicted that the strong-coupling and retardation effects play pivotal role in the superconducting phase of HTe at 300 GPa, in agreement with similar theoretical estimates for the sibling hydrogen and hydrogen-based compounds. Hence, it is suggested that the superconducting state in HTe cannot be properly described within the mean-field picture of the BCS theory.

\end{abstract}

\maketitle


\section{Introduction}

Recent theoretical and experimental investigations prove that chalcogen atoms (particularly elements of group VIa) paired with hydrogen constitute promising compounds for applications as a phonon-mediated high-temperature superconductors \cite{drozdov1, durajski, drozdov2, szczesniak1, kim, eremets}. The representative example of such compounds is hydrogen sulfide (H$_{3}$S), which was theoretically and experimentally proved to exhibit record high superconducting critical temperature value ($T_{c}$) of 203 K at megabar pressure \cite{drozdov1, durajski}. The origin of the idea of inducing high-temperature superconducting phase in hydrogen-based materials steams from the milestone work of Ashcroft for the metallic hydrogen \cite{ashcroft1}. Much later, in 2004, the same author suggested that adding heavier elements into the metallic hydrogen leads to the chemical pre-compression of the resulting compound, which lowers pressure value required for the induction of the superconducting phase while still retaining relatively high values of $T_{c}$ \cite{ashcroft2}.

Another promising candidate for the applications described above is tellurium, which is isostructural with sulfur \cite{hejny} while having larger atomic core and weaker electronegativity \cite{zhong}. The recent density functional theory simulations show that it is possible to induce phonon-mediated superconducting phase in HTe, with $T_{c}$ reaching maximum value at 300 GPa \cite{zhong}. In fact, the pressure diagram in HTe share similar features with H$_{3}$S {\it i.e.} first the critical temperature increases with pressure, to be followed by the sudden strong decrease. Further, another increase in $T_{c}$ is observed with increasing pressure to reach mentioned maximum at 300 GPa in the case of HTe \cite{zhong}. The main difference between phase diagrams of HTe and H$_{3}$S is the fact that for the latter compound the highest critical temperature value is obtained at the first maxima around 150 GPa \cite{durajski}.

For the further development in the discussed field, it is natural that novel candidate materials, such as HTe, has to be understood in details. In this context, herein the complementary and comprehensive investigations of the superconducting phase are provided for  the HTe at 300 GPa; the pressure conditions which are considered to be a perfect case study and match the state where superconducting phase in HTe reaches its maximum value of $T_{c}$. Due to the relatively high value of the electron-phonon coupling constant ($\lambda = 1.07$) in HTe at 300 GPa, the discussed superconductor is investigated within the strong-coupling generalization of the Bardeen-Cooper-Schrieffer (BCS) theory, namely the formalism of the Eliashberg equations \cite{eliashberg}; note that such strategy is suggested for the materials characterized by $\lambda>0.5$ \cite{carbotte, cyrot}. Moreover, the isotropic form of the Fermi surface in HTe \cite{zhong} reinforces the choice of corresponding form of the Eliashberg equations. Specifically, the present paper is aimed at the determination of the thermodynamic properties of the superconducting state in HTe, such as the critical temperature value, the energy band gap, the effective mass of electrons, the thermodynamic critical field, and the specific heat. In addition to the detailed characterization, the present study provides vital test of the strength of the electron-phonon interactions, and their role in the analyzed superconducting state.

\section{Results and Discussion}

%
\begin{figure}[ht!]
\centering
\includegraphics[width=\columnwidth]{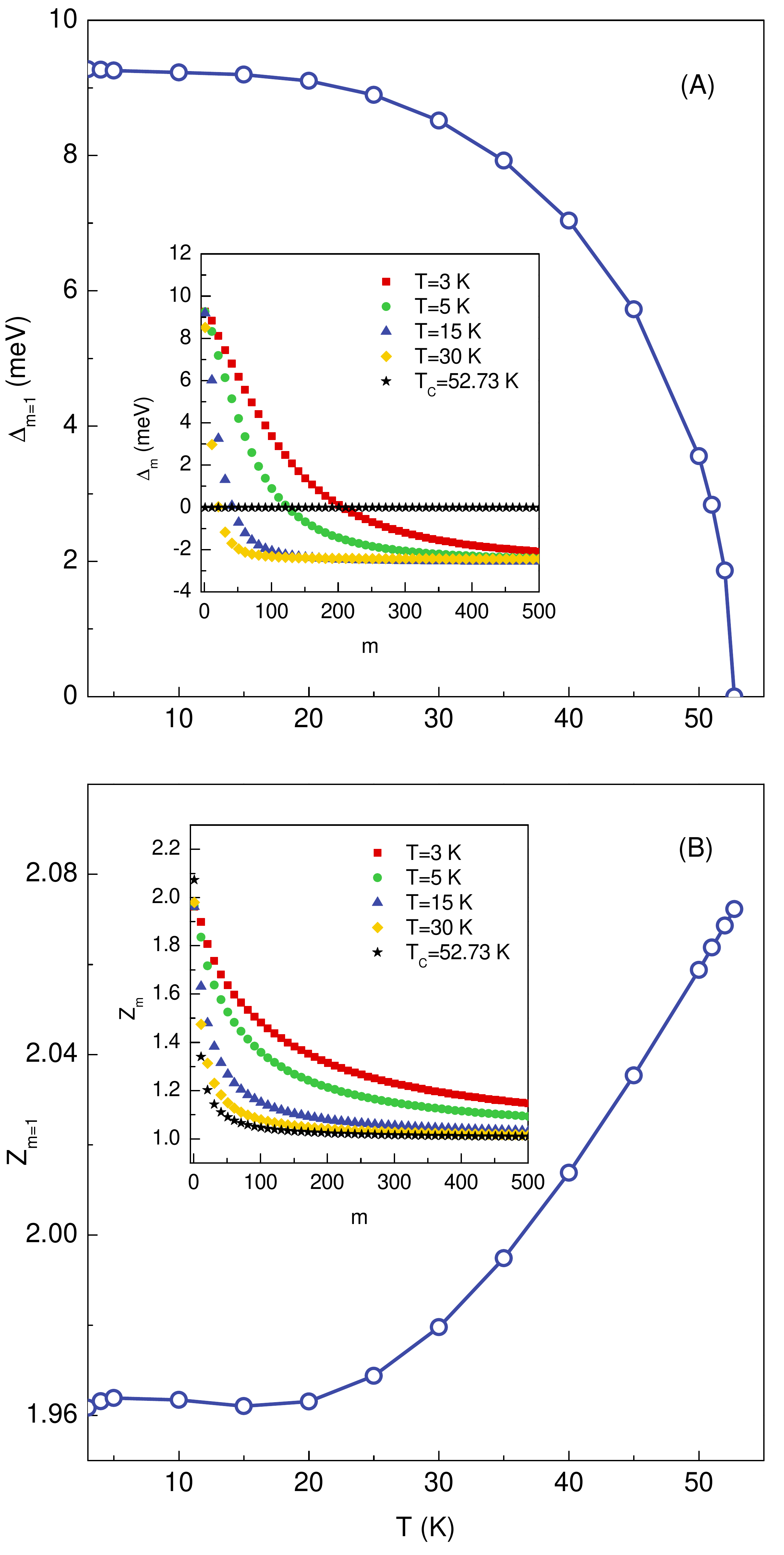}
\caption{(A) The maximum value of the order parameter ($\Delta_{m=1}$) and (B) the wave-function renormalization factor ($Z_{m=1}$) as a function of the temperature ($T$), for HTe at 300 GPa. Insets given in (A) and (B) show the thermal behavior of the order parameter ($\Delta_{m}$) and the wave-function renormalization factor ($Z_{m}$) on $m$ for the selected values of the temperature, for the analyzed superconductor.}
\label{fig1}
\end{figure}

As already mentioned, in the present study the thermodynamics of the superconducting state in the tellurium hydride (HTe) at very high pressure ($p=300$ GPa) are determined in the framework of the Eliashberg equations \cite{eliashberg}. This model is a natural generalization of the Bardeen-Cooper-Schrieffer theory \cite{bardeen1, bardeen2}, and is adopted here due to the relatively high value of the electron-phonon coupling constant ($\lambda=1.07$) in HTe at 300 GPa. Specifically, the Eliashberg formalism is employed in the isotopic form within the Migdal approximation \cite{migdal}, in correspondence to the isotropic approximation of pairing gap proposed in \cite{zhong}. Herein, the Eliashberg equations are solved on the imaginary axis, and later analytically continued on the real axis, by using the numerical methods employed previously in \cite{szczesniak2, szczesniak3, szczesniak4, szczesniak5, szczesniak6}. The following form of the Eliashberg equations on the imaginary axis ($i\equiv\sqrt{-1}$) is adopted during calculations:
\begin{equation}
\label{eq1}
\Delta_{n}Z_{n}=\frac{\pi}{\beta} \sum_{m=-M}^{M}
\frac{K\left(\omega_{n}-\omega_{m}\right)-\mu^{\star}\left(\omega_{m}\right)} 
{\sqrt{\omega_m^2+\Delta_m^2}}
\Delta_{m}, 
\end{equation}
and
\begin{equation}
\label{eq2}
Z_n=1+\frac {\pi}{\beta\omega _n }\sum_{m=-M}^{M}
\frac{K\left(\omega_{n}-\omega_{m}\right)}
{\sqrt{\omega_m^2+\Delta_m^2}}\omega_m.
\end{equation}
In this context, above set of equations allow to find the values of the order parameter ($\Delta_{n}\equiv\Delta\left(i\omega_{n}\right)$) and the wave function renormalization factor ($Z_{n}\equiv Z\left(i\omega_{n}\right)$), respectively. Note that, for the Eliashberg equations on the imaginary axis, the frequency takes on the discrete values ($\omega_{n}\equiv \frac{\pi}{\beta}\left(2n-1\right)$) which are temperature dependent ($\beta\equiv 1/k_{B}T$, where $k_{B}$ is the Boltzmann constant). Herein, the numerical stability is ensured by taking into account 2201 Matsubara frequencies. In this manner, the high-accuracy predictions are obtained for $T \geq T_{0}$, where $T_{0}=3$ K, assuming that the cut-off frequency $\omega_{c}=10\Omega_{\rm max}$, where $\Omega_{\rm max}$ is the maximum phonon frequency equal to $188.59$ meV. Moreover, in Eqs. (\ref{eq1}) and (\ref{eq2}), the $K\left(\omega_{n}-\omega_{m}\right)$ function denotes the pairing kernel for the electron-phonon interaction: $K\left(\omega_{n}-\omega_{m}\right)\equiv 2\int_0^{\Omega_{\rm{max}}}d\Omega\frac{\alpha^{2}F\left(\Omega\right)\Omega}{\left(\omega_n-\omega_m\right)^2+\Omega ^2}$. Next, the quantity $\alpha^{2}F\left(\Omega\right)$ is the so-called Eliashberg function, which models the electron-phonon interactions in the considered material. For the tellurium hydride at the $p=300$ GPa, the form of the $\alpha^{2}F\left(\Omega\right)$ function is adopted after \cite{zhong}; in the referred paper $\alpha^{2}F\left(\Omega\right)$ function was calculated within the linear response theory via the Quantum-ESPRESSO computational suite \cite{giannozzi}.

In the present analysis, the depairing electronic correlations are modeled by the phenomenological parameter known as the Coulomb pseudopotential ($\mu^{\star}$), which assumes value of 0.1. The magnitude of $\mu^{\star}$ is adopted to match predictions of Ashcroft for the hydrogen-based materials \cite{ashcroft2} and to treat our results on the same footing with the analytical results for the HTe at 300 GPa, as presented in \cite{zhong}.

In Fig. \ref{fig1} (A) and (B), the solutions of the Eliashberg equations on the imaginary axis are presented. In particular, Fig. \ref{fig1} (A) depicts the temperature dependance of the order parameter function, whereas Fig. \ref{fig1} (B) gives the functional behavior of the wave function renormalization factor on the temperature. For the both presented functions results are plotted for $T \in < T_{0}, T_{c} >$, where $T_{c}$ denotes the critical value of the temperature. In Fig. \ref{fig1} (A), the maximum value of the order parameter ($\Delta_{m=1}\left(T\right)$) allows to determine the $T_{c}$ by using the following relation: $\Delta_{m=1}\left(T_{c}\right)=0$. In this context, the determined $T_{c}$ value for the HTe at 300 GPa is equal to 52.73 K. Note that the McMillan analytical model \cite{mcmillan, allen}, as presented in \cite{zhong}, gives $T_{c}=44.26$ K; therefore more than 8 K lower than the estimate of the Eliashberg formalism. It is the first indication that the Eliashberg equations may provide paramount corrections to the value of the pivotal $T_{c}$ quantity. In particular, noted discrepancies may be caused by the strong-coupling and retardation effects which governs superconducting phase in HTe at 300 GPa, and are included within the Eliashberg formalism which is not the case for McMillan analytical model based on the BCS theory. Note, that in our opinion the obtained $T_{c}$ is not influenced by the insufficient accuracy of the conducted numerical calculations. In fact, assumed number of the Matsubara frequencies (equals to 2201) is well above the saturation point of the imaginary axis solutions for the order parameter function; see inset of Fig. \ref{fig1} (A), which presents the order parameter ($\Delta_{m}$) as a function of $m$ for the selected values of temperature, where saturation point for the $\Delta_{m}$ can be observed for $m \sim 400$ at all selected temperatures.

To further investigate the character of the superconducting phase in the HTe at 300 GPa, the behavior of the wave function renormalization factor is investigated. As already mentioned, Fig. \ref{fig1} (B) presents the thermal behavior of the maximum value of the wave function renormalization factor ($Z_{m=1}\left(T\right)$). Note that similarly to the order parameter function, the $Z_{m}$ function also presents saturation of the Eliashberg equations solutions for $m \sim 400$, as depicted in the inset of the Fig. \ref{fig1} (B). From the physical point of view, the wave function renormalization factor with a good accuracy reproduces the thermal evolution of the effective mass of electrons ($m^{\star}_{e}$), according to the following relation: $m^{\star}_{e}\simeq Z_{m=1}(T) m_{e}$, where $m_{e}$ denotes the band electron mass. In this context, the effective mass of electrons in the HTe at 300 GPa is equal to 2.07 $m_{e}$. This value is in perfect agreement with the predictions from the fundamental analytical relation for the phonon-mediated superconductors {\it i.e.} $\left[m^{\star}_{e}/m_{e}\right]_{T=T_{c}}\simeq 1+\lambda\simeq 2.07$ \cite{carbotte}. Note however, that both values are way above limit set by the BCS theory which states that $m^{\star}_{e}=m_{e}$ at $T=T_{c}$ \cite{bardeen1, bardeen2}. Therefore, the second parameter, after calculated above $T_{c}$ value, suggests important role of the strong-coupling effects in the HTe at 300 GPa.
 
\begin{figure}[ht]
\centering
\includegraphics[width=\columnwidth]{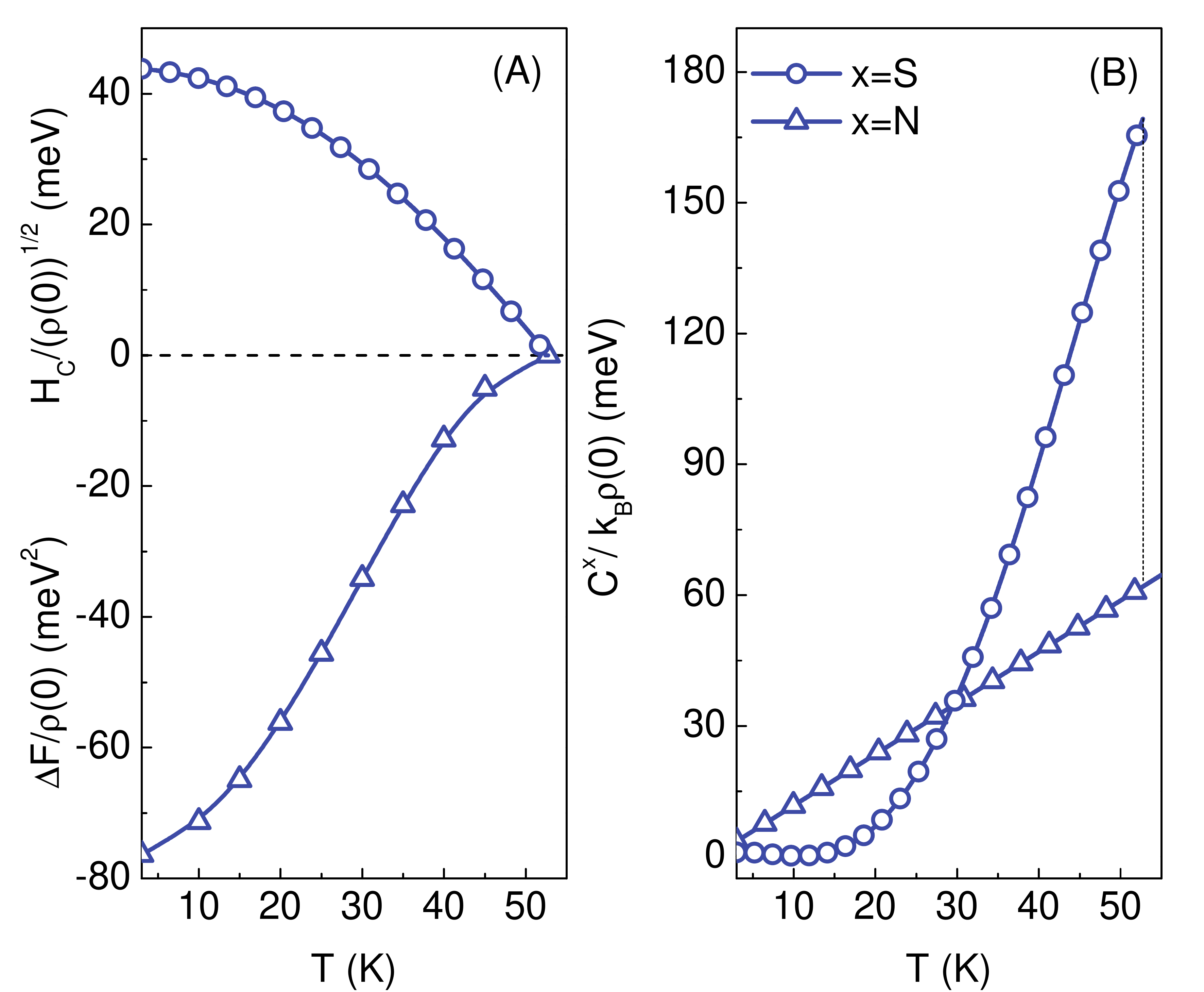}
\caption{(A) The normalized critical thermodynamic field ($H_{C}/\sqrt{\rho\left(0\right)}$) (upper panel) and the normalized difference of the free energy for the normal and superconducting state ($\Delta F/\rho(0)$) (lower panel) as a function of the temperature, for HTe at 300 GPa. The thermal behavior of the normalized specific heat for the superconducting ($C^{S}\left(T\right)/k_{B}\rho\left(0\right)$) and the normal state ($C^{N}\left(T\right)/k_{B}\rho\left(0\right)$) on the temperature, for the analyzed superconductor.}
\label{fig2}
\end{figure}

To further verify the strong-coupling character of the superconducting state in the analyzed hydrogen-based material, it is instructive to calculate the characteristic dimensionless ratios for the thermodynamic critical field ($H_{c}$), the specific heat for the superconducting state ($C^{S}$), and the superconducting band gap ($\Delta_{g}$), towards their comparison with  the predictions of the BCS theory. Specifically, the thermodynamic critical field normalized with respect to the density of states at the Fermi level ($\rho\left(0\right)$), can be written as \cite{blezius, carbotte}:
\begin{eqnarray}
\label{eq3}
\frac{H_{C}}{\sqrt{\rho\left(0\right)}}=\sqrt{-8\pi\frac{\Delta F}{\rho\left(0\right)}},
\end{eqnarray}  
where $\Delta F/\rho\left(0\right)$ denotes the normalized free energy difference between the superconducting and normal state, given as \cite{bardeen3}:
\begin{eqnarray}
\label{eq4}
\frac{\Delta F}{\rho\left(0\right)}&=&-\frac{2\pi}{\beta}\sum_{m=1}^{M}
\left(\sqrt{\omega^{2}_{m}+\Delta^{2}_{m}}- \left|\omega_{m}\right|\right)\\ \nonumber
&\times&(Z^{{S}}_{m}-Z^{N}_{m}\frac{\left|\omega_{m}\right|}{\sqrt{\omega^{2}_{m}+\Delta^{2}_{m}}}).
\end{eqnarray}  
In Eq. (\ref{eq4}), the $Z^{S}_{m}$ and $Z^{N}_{m}$ denote the wave function renormalization factor for the superconducting state ($S$) and the normal state ($N$), respectively. The determined thermal dependence of the functions given by Eqs. (\ref{eq3}) and (\ref{eq4}) are depicted in the upper and lower panel of Fig. \ref{fig2} (A), respectively.

Note that the $H_{C}/\sqrt{\rho\left(T\right)}$ function takes on only positive values and decreases together with the increase of the temperature, where above $\sim 30$ K mentioned decrease is almost linear. The maximum value of $H_{C}/\sqrt{\rho\left(T\right)}$ is equal to 44.09 meV at $T=T_{0}$. In overall, such characteristics is in agreement with the behavior presented by the phonon-mediated superconductors \cite{carbotte}. On the contrary, the $\Delta F/\rho(0)$ presents only negative values, meaning that the obtained results satisfy the third law of the thermodynamics.

The knowledge of the free energy allows also to calculate the normalized specific heat difference between the superconducting and normal state ($\Delta C/k_{B}\rho\left(0\right)$) as \cite{blezius, carbotte}: 
\begin{eqnarray}
\label{eq5}
\Delta C/k_{B}\rho\left(0\right)=-\frac{1}{\beta}\frac{d^{2}\left[\Delta F/\rho\left(0\right)\right]}{d\left(k_{B}T\right)^{2}}.
\end{eqnarray}  
In the context of Eq. (\ref{eq5}), the specific heat for the superconducting state ($C^S$) is derived from the following relation: $\Delta C\equiv C^S-C^N$, where the specific heat in the normal state should be calculated as: $C^{N}/k_{B}\rho\left(0\right)=\gamma/\beta$, assuming $\gamma\equiv\frac{2}{3}\pi^{2}\left(1+\lambda\right)$. The full functional dependence of the $C^S$ and $C^N$ functions, normalized with respect to the $\rho\left(0\right)$, is plotted in Fig. \ref{fig2} (B). The difference between both mentioned functions is cleary visible, where the specific heat for the superconducting state at $T_{c}$ presents characteristic {\it jump}.

Next, the analysis conducted above makes it possible to determine two of the already mentioned characteristic dimensionless ratios {\it i.e}: $R_{H}\equiv T_{c}C^{N}\left(T_{c}\right)/H^{2}_{C}\left(0\right)$ and $R_{C}\equiv \Delta C\left(T_{c}\right)/C^{N}\left(T_{c}\right)$. The computed values of these ratios for the HTe at 300 GPa are: $R_{H}=0.147$ and $R_{C}=1.73$. Note that in the framework of the BCS theory both parameters adopt the following universal values: $\left[R_{H}\right]_{\rm BCS}=0.168$ and $\left[R_{C}\right]_{\rm BCS}=1.43$ \cite{bardeen1, bardeen2}. Therefore, estimates presented herein visibly exceeds prediction of the BCS theory, once again suggesting paramount role of the strong-coupling and retardation effects in the analyzed material.

The third and the last characteristic ratio of interest is calculated for the value of the superconducting energy band gap at the Fermi level which is defined as:  $\Delta_{g} \equiv 2\Delta\left(0\right)$. Herein, the exact $\Delta\left(0\right)$ value is estimated by analytical continuation of the imaginary axis solutions on the real axis. In particular, the order parameter function, within the employed here Pad{\' e} method \cite{beach}, is given as: $\Delta\left(T\right)={\rm Re}\left[\Delta\left(\omega=\Delta\left(T\right),T\right)\right]$, where $\Delta\left(\omega\right)$ is calculated from the following relation:
\begin{equation}
\label{eq6}
\Delta\left(\omega\right)=\frac{p_{1}+p_{2}\omega+...+p_{r}\omega^{r-1}}
{q_{1}+q_{2}\omega+...+q_{r}\omega^{r-1}+\omega^{r}}.
\end{equation}
In Eq. (\ref{eq6}), the $p_{j}$ and $q_{j}$ are the number coefficients and $r=500$. 

\begin{figure}[ht]
\centering
\includegraphics[width=\columnwidth]{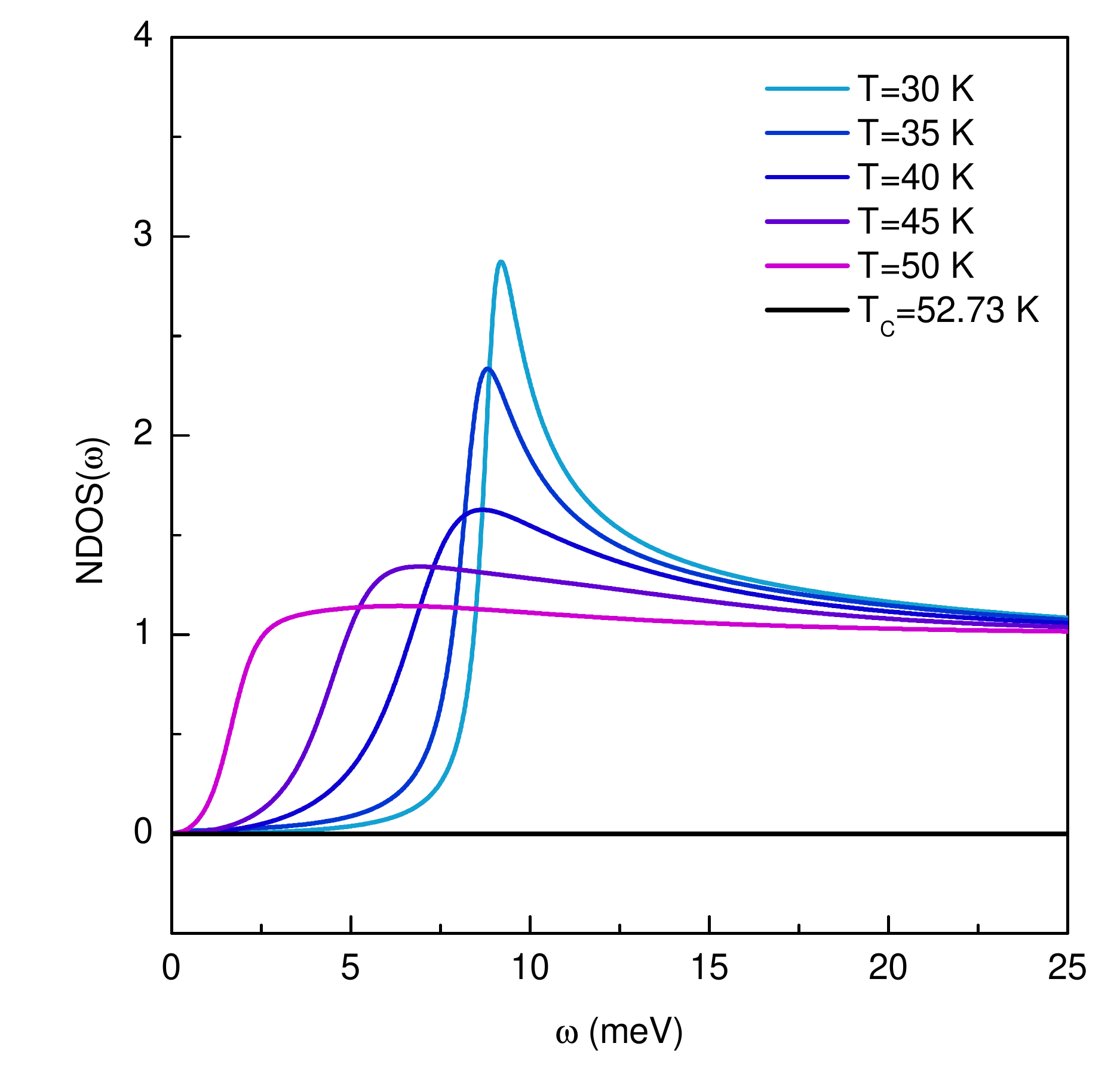}
\caption{The normalized density of states as a function of frequency (NDOS($\omega$)) for the selected temperature values in HTe at 300 GPa.}
\label{fig3}
\end{figure}

The thermal dependence of the calculated paring gap is presented in Fig. \ref{fig3}, in a form of the normalized electronic density of states (NDOS) given as:
\begin{equation}
\label{eq7}
{\rm NDOS}\left(\omega \right)=\frac{\rm DOS_{S}\left(\omega \right)}{\rm DOS_{N}\left(\omega \right)}={\rm Re}\left[\frac{\left|\omega -i\Gamma \right|}{\sqrt{\left(\omega -i\Gamma\right)^{2}}-\Delta^{2}\left(\omega\right)}\right],
\end{equation}
where $\rm DOS_{S}\left(\omega \right)$ and $\rm DOS_{N}\left(\omega \right)$ is the density of states for the superconducting and normal state, respectively, whereas $\Gamma$ denotes the pair breaking parameter equal to $0.15$ meV.

As depicted in Fig. \ref{fig3}, together with the temperature decrease the magnitude of pairing gap increases. In this context, the most representative value of the pairing gap is given at T=0 K. Therefore, the determined exact value of the energy gap at the Fermi level for the HTe at 300 GPa is equal to $2\Delta\left(0\right)=18.93$ meV. Then, the last remaining dimensionless ratio can be calculated as: $R_{\Delta}\equiv 2\Delta\left(0\right)/k_{B}T_{c}$, and its value is equal to $4.18$. Again, determined value of the characteristic dimensionless ratio exceeds predictions of the classical BCS theory, which states that $\left[R_{\Delta}\right]_{\rm BCS}=3.53$ \cite{bardeen1, bardeen2}. Thus, similarly to the $R_{H}$ and $R_{C}$ ratios, the obtained above ratio strongly deviate from the BCS model.

\section{Summary and conclusions}

In the present paper, the superconducting state in the HTe at 300 GPa was analyzed within the Eliashberg formalism towards its complementary characterization and to prove the strong coupling between electrons and phonons in this phase. In particular, the value of the critical temperature was estimated to be equal to $T_{c}=52.73$ K, adopting universal value of the Coulomb pseudopotential for the hydrogen-based materials equal to $\mu^{*}=0.1$ \cite{ashcroft2}. Determined value appeared to be almost 8 K higher than the estimate of the McMillan expression, which base on the BCS theory \cite{zhong}. Observed discrepancy was attributed to the strong-coupling and retardation effects which may be present in the analyzed superconducting phase. This suggestion was further verified by the calculation of the characteristic dimensionless ratios, familiar in the BCS theory; the characteristic dimensionless ratios for the thermodynamic critical field ($R_{H}$), the specific heat for the superconducting state ($R_{C}$), and the superconducting band gap ($R_{\Delta}$). The obtained values of the $R_{H}$, $R_{C}$, and $R_{\Delta}$ ratios equal to 0.147, 1.73 and 4.17, respectively. Hence, all of the discussed parameters visibly exceeds predictions of the BCS theory, where $\left[R_{H}\right]_{\rm BCS}=0.168$, $\left[R_{C}\right]_{\rm BCS}=1.43$, and $\left[R_{\Delta}\right]_{\rm BCS}=3.53$ \cite{bardeen1, bardeen2}. Such behavior can be considered as a vital test proving that the superconducting phase in HTe at 300 GPa is strongly governed by the many-body effects (with pivotal role of the strong-coupling and retardation effects), and cannot be properly characterized within the mean-field BCS theory picture. Note that such characteristics are in agreement with recent findings for other sibling hydrogen \cite{szczesniak7, szczesniak10} and hydrogen-based materials \cite{durajski, duda, szczesniak1, szczesniak8, szczesniak9} analyzed within the Eliashberg formalism. This observation is additionally reinforced by the relatively high value of the effective electron mass ($m^{\star}_{e}$), which is not equal to the bare electron mass ($m_{e}$) at $T=T_{c}$, as predicted by the BCS theory \cite{bardeen1, bardeen2}.

\begin{acknowledgments}
The Authors would like to note that partial numerical calculations were conducted on the Cz{\c e}stochowa University of Technology cluster, built in the framework of the PLATON project no. POIG.02.03.00-00-028/08 - the service of the campus calculations U3.
\end{acknowledgments}
\bibliographystyle{apsrev}
\bibliography{bibliography}
\end{document}